\documentclass[aps,prb,amsmath,amssymb,reprint,superscriptaddress,preprintnumbers,showpacs,intlimits,longbibliography]{revtex4-1}
\pdfoutput=1
\bibpunct{[}{]}{,}{n}{}{}

\usepackage{bm,latexsym,mathrsfs,enumerate}
\usepackage{xcolor}
\usepackage{mathtools}
\usepackage[mathcal]{euscript}
\usepackage[breaklinks=true,unicode=true,urlcolor = blue,colorlinks = true,citecolor = blue,linkcolor = blue]{hyperref}
\usepackage{pgf,tikz}
\usepackage{graphicx}
\usepackage{todonotes}
\usepackage{wasysym}
\usepackage{media9}
\renewcommand{\vec}[1]{\bm{#1}}
\usepackage[normalem]{ulem}

\begin{document}

\title{Curvature and torsion effects in the spin-current driven domain wall motion}

\author{Kostiantyn V. Yershov}
\email{yershov@bitp.kiev.ua}
\affiliation{Bogolyubov Institute for Theoretical Physics of National Academy of Sciences of Ukraine, 03680 Kyiv, Ukraine}
\affiliation{National University of ``Kyiv-Mohyla Academy", 04655 Kyiv, Ukraine}

\author{Volodymyr P. Kravchuk}
\email{vkravchuk@bitp.kiev.ua}
\affiliation{Bogolyubov Institute for Theoretical Physics of National Academy of Sciences of Ukraine, 03680 Kyiv, Ukraine}

\author{Denis D. Sheka}
\email{sheka@univ.net.ua}
\affiliation{Taras Shevchenko National University of Kyiv, 01601 Kyiv, Ukraine}

\author{Yuri Gaididei}
\email{ybg@bitp.kiev.ua}
\affiliation{Bogolyubov Institute for Theoretical Physics of National Academy of Sciences of Ukraine, 03680 Kyiv, Ukraine}

\date{November 6, 2015}

%
%
%
%
\begin{abstract}
The domain wall motion along a helix-shaped nanowire is studied for the case of spin-current driving via Bazaliy--Zhang--Li mechanism. The analysis is based on collective variable approach. Two new effects are ascertained: (i) the curvature results in appearance of the Walker limit for a uniaxial wire, (ii) the torsion results in effective shift of the nonadiabatic spin torque parameter $\beta$. The latter effect changes considerably the domain wall velocity and can result in negative domain wall mobility. This effect can be also used for an experimental determination of the nonadiabatic parameter $\beta$ and damping coefficient $\alpha$.
\end{abstract}

\maketitle

\section{Introduction}

The Walker limit \cite{Schryer74,Thiaville06,Mougin07} is a well known property of a one dimensional domain wall motion in a biaxial magnet. This phenomenon establishes existence of a critical value $f_c$ of a driving force (e.g. applied magnetic field \cite{Schryer74,Bouzidi90,Thiaville06,Mougin07,Goussev13} or spin-polarized current \cite{Thiaville05,Mougin07,Fukami08,Goussev13}), which corresponds to the bifurcation \cite{Goussev13} between two different regime of the domain wall motion: traveling-wave motion for $f<f_c$ and precession regime for $f>f_c$. The precession regime is characterized by precession of the domain wall magnetization around the wire. In this regime the translation motion of the domain wall along the wire is oscillating one for a biaxial magnet  \cite{Thiaville06}, however in the limit case of the uniaxial magnet the translational motion becomes stationary~ \cite{Yan10}. Transition to the precession regime is usually characterized by the rapid decrease (breakdown) of the averaged domain wall velocity (however, the rapid increase is also possible, e.g. for the case of spin-current driving with small nonadiabaticity \cite{Thiaville04,Tatara04,Thiaville05}).  

The critical value $f_c$ is linearly proportional to the coefficient of the transversal anisotropy \cite{Schryer74,Thiaville06} and, therefore $f_c=0$ for an uniaxial magnet, it means that the traveling-wave motion is not possible in this case and the precession regime appears for any value of $f>0$. The \emph{first aim} of the current study is to demonstrate that the domain wall motion in uniaxial curvilinear wire possesses the bifurcation picture with the critical value $f_c\propto\kappa>0$ with $\kappa$ being the geometrical curvature of the wire. This effect originates from the curvature induced effective Dzyaloshinskii--Moria interaction (DMI) \cite{Sheka15}.

For the case of spin-current driving via Bazaliy--Zhang--Li mechanism the parameter of the spin-torque nonadiabaticity \cite{Zhang04} $\beta$ is fundamentally important. It is well known \cite{Tatara04,Thiaville05,Mougin07,Fukami08,Goussev13,Zhang04} that for the case $\beta=0$ a domain wall does not move \footnote{At the initial moment of the current application the domain wall can demonstrate a short shift, however finally it stops and no steady motion is observed.} when $f<f_c$, i.e. the traveling-wave motion is not possible. The \emph{second aim} of the current study is to demonstrate that for the case of a curvilinear wire the nonadiabaticity parameter effectively experiences a geometrically induced shift $\beta\to\beta-\beta^\star$ with $\beta^\star\propto\tau$ with $\tau$ being the geometrical torsion of the wire. This effect originates from torsion induced effective DMI \cite{Sheka15}. It can result in negative effective nonadiabaticity parameter and in this case the domain wall demonstrates negative mobility, i.e. it moves against the electrons flow. In a particular purely adiabatic case $\beta=0$ domain walls demonstrate nonzero mobilities, moreover the mobility sign is determined by product of the helix chirality and domain wall topological charge (head-to-head or tail-to-tail). These phenomena are not observed for a rectilinear wire. However, for a helix nanowire the similar behavior of the mobility was recently demonstrated for the Rashba torque driven domain wall motion \cite{Pylypovskyi15e}. In some respect, the effect of chirality sensitive domain wall mobility is similar to the recently found chiral-induced spin selectivity effect in helical molecules \cite{Goehler11,Naaman12,Eremko13}.

The paper is organized as follows: in Section~\ref{sec:geometry} we introduce basic geometrical notations and parameterization of a helix wire; in Section~\ref{sec:eqs-of-motion} we demonstrate how the basic equations of motion with the corresponding Lagrange formalization are modified due to the curvature and torsion; in Section~\ref{sec:collective-variables} we apply the collective variable approach in order to study static and dynamic properties of the domain wall; main results are summarized in Section~\ref{sec:concl}; in Appendix~\ref{app:CV-equations} the details of the collective variable approach are presented; bifurcation analysis of the domain wall dynamics is performed in Appendix~\ref{app:bifurcation}; the supplemental movies are presented in Appendix~\ref{app:movie}.

\section{Formalization of geometry}\label{sec:geometry}
To determine the role of curvature and torsion in the domain wall dynamics we consider a wire in form of three dimensional helix as a case study. In this case the geometrical effects are the most demonstrable due to constant curvature and torsion. We parameterize the helix curve $\vec\gamma(s)$ in the following way
\begin{equation} \label{eq:helix-simple}
\vec{\gamma}(s)=\mathcal{R}\left[\vec{e}_x \cos\frac{2\pi s}{s_0}+\vec{e}_y\sin\frac{2\pi s}{s_0}\right]+\vec{e}_z\mathcal{C}\frac{s}{s_0}\mathcal{P}.
\end{equation}
where $s$ is natural parameter (arc length), $\left(\vec{e}_x,\vec{e}_y ,\vec{e}_z\right)$ is Cartesian basis, $\mathcal{R}$ and $\mathcal{P}$ are radius and pitch of the helix, respectively, and $\mathcal{C}=\pm1$ is the helix chirality, right ($\mathcal{C}=+1$) or left ($\mathcal{C}=-1$). Parameter $s_0=\sqrt{\mathcal{P}^2 + 4\pi^2 \mathcal{R}^2}$ is length of a single helix coil. However, the Cartesian reference frame is not convenient for a curvilinear wire, therefore in the following we proceed to local Frenet--Serret reference frame, described by the curvelinear basis vectors ($\vec{e}_\textsc{{t}},\,\vec{e}_\textsc{{n}},\vec{e}_\textsc{{b}}$) with $\vec{e}_\textsc{{t}}=\vec{\gamma}'(s)$, $\vec{e}_\textsc{{n}}=\vec{\gamma}''(s)/|\vec{\gamma}''(s)|$, and $\vec{e}_\textsc{{b}}=\vec{e}_\textsc{{t}}\times\vec{e}_\textsc{{n}}$ being the tangential, normal, and binormal unit vectors, respectively. Here and below the prime denotes the derivative with respect to the arc length coordinate $s$. In contrast to the Cartesian basis the local basis is spatially dependent and its differential properties are determined by the Frenet--Serret formulas
\begin{equation} \label{eq:FS-formulas}
	\vec{e}_\textsc{t}'(s)=\kappa\vec{e}_\textsc{n},\,\;\;\vec{e}_\textsc{n}'(s)=-\kappa\vec{e}_\textsc{t}+\tau\vec{e}_\textsc{b},\,\;\; \vec{e}_\textsc{b}'(s)=-\tau\vec{e}_\textsc{n}.
\end{equation}
One can consider \eqref{eq:FS-formulas} as a definition of curvature $\kappa$ and torsion $\tau$ of the wire. A helix curve has a specific feature: the curvature $\kappa=4\pi^2 \mathcal{R}/s_0^2$ as well as the torsion $\tau = 2\pi \mathcal{C}\mathcal{P}/s_0^2$ are constants. By setting values of both parameters $\kappa$ and $\tau$ (with sign) one determines the helix curve in unique way, thus, in the following discussion we use $\kappa$ and $\tau$ as only geometrical parameters. 

\begin{figure*}[t]
 	\includegraphics[width=\textwidth]{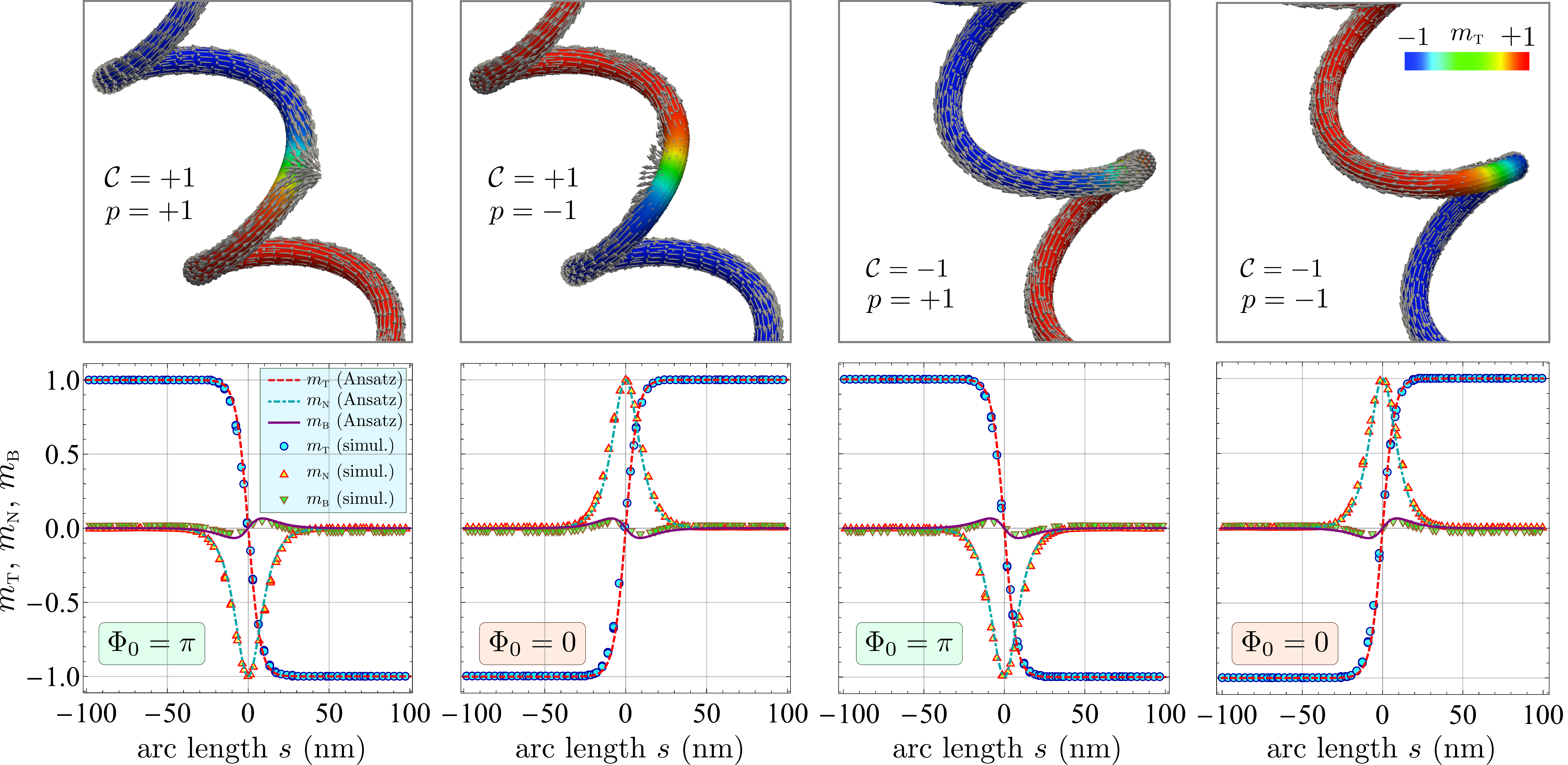}
 	\caption{\label{fig:stat-state}%
 		(Color online) Static domain walls on the helix wire. Top row shows the magnetization distribution obtained using micromagnetic \texttt{NMAG} simulations for permalloy helix wire with $\rho_0 = 5$, total length $L = 6$ $\mu$m (only vicinities of the domain wall positions are shown). Helix radius and pitch are $\mathcal{R}=30$~nm and $\mathcal{P}=93$~nm, respectively ($\kappa\Delta_0=0.2$, $|\tau\Delta_0|=0.1$). The bottom row demonstrates comparison of the simulation data (markers) and Ansatz \eqref{eq:q-Phi-gen} (lines), with the domain wall width $\Delta=\Delta_0=2\ell$ (magnetically soft material). }
 \end{figure*}

In order to consider a physical wire of finite thickness, one can use the following  parameterization
 \begin{equation} \label{eq:wire-shape}
 	\vec{r}(s,\chi,\rho)=\vec{\gamma}(s)+\rho\cos\chi\vec{e}_\textsc{{n}}(s)+\rho\sin\chi\vec{e}_\textsc{{b}}(s).
 \end{equation}
 Here the three-dimensional radius vector $\vec{r}$ defines the space domain, occupied by the wire, $\rho\in[0,\,\rho_0]$ and $\chi\in[0,\,2\pi)$ are coordinates within the wire cross-section with $\rho_0$ being the wire radius. For examples of a helix wire defined by \eqref{eq:wire-shape} and \eqref{eq:helix-simple} see Fig.~\ref{fig:stat-state}. 
 
\section{Equations of motion}\label{sec:eqs-of-motion}
By reasons of the nontrivial curvilinear geometry the spin polarized current is more preferable driving for a domain wall on the helix wire as compared with the external magnetic field. Thus, we base our study on the Landau--Lifshitz equation with additional Bazaliy--Zhang--Li spin-torque terms \cite{Bazaliy98,Zhang04,Thiaville05}
\begin{equation} \label{eq:LL-ZhLi}
\begin{split}
\dot{\vec{m}}=&\;\omega_0\vec{m}\times\frac{\delta\mathcal{E}}{\delta\vec{m}} + \alpha\vec{m}\times\dot{\vec{m}}\\
+&\vec{m} \times \left[\vec{m}\times\left(\vec{u}\cdot\nabla\right) \vec{m}\right] + \beta\vec{m} \times \left(\vec{u}\cdot\nabla\right)\vec{m}.
\end{split}
\end{equation}

Here $\vec{m}=\vec{M}/M_s$ is the unit magnetization vector with $M_s$ being the saturation magnetization. The overdot indicates derivative with respect to time and the characteristic timescale of the system is determined by $\omega_0=4\pi\gamma_0M_s$ with $\gamma_0$ being the gyromagnetic ratio. Here $\mathcal{E}=E/(4\pi M_s^2)$ is normalized total energy of the system. The driving strength is presented by the quantity $\vec{u}=\vec{j}P\mu_B/(|e|M_s)$ which is close to average electron drift velocity in presence of the current of density $\vec{j}||\vec{e}_\textsc{t}$ \cite{Tserkovnyak08}, here $P$ is the rate o spin polarization, $\mu_B$ is Bohr magneton, and $e$ is electron charge. Constants $\alpha$ and $\beta$ denote Gilbert damping and the nonadiabatic spin-transfer parameter, respectively. For detailed derivation of spin-torques and the applications see reviews Refs.~\onlinecite{Ralph08a,Tserkovnyak08,Tatara08,Brataas12}.

To write the energy functional $\mathcal{E}$ we consider a simple magnetic wire model, which takes into account only two contributions to the total magnetic energy
\begin{equation} \label{eq:total-energy}
\mathcal{E} =\mathcal{S}\int_{-\infty}^{+\infty} \bigl[\ell^2\mathscr{E}_{\text{ex}}-k_{\text{t}}(\vec{m}\cdot\vec{e}_\textsc{t})^2\bigr]\text{d}s,
\end{equation}
namely, exchange one $\mathscr{E}_{\text{ex}}$ and easy-tangential anisotropy -- the second term in \eqref{eq:total-energy}. Here $\mathcal{S}=\pi\rho_0^2$ is area of the wire cross-section, $\ell=\sqrt{A/4\pi M_s^2}$ is the exchange length with $A$ being exchange constant, and $k_{\text{t}}=K/(4\pi M_s^2)+1/4$ is the dimensionless anisotropy constant. Here $K>0$ is an easy-tangential magnetocrystalline anisotropy constant and the term $1/4$ comes from the magnetostatic contribution  \cite{Thiaville06,Porter04,Slastikov12,Kravchuk14c}.  Recently we demonstrated that the model \eqref{eq:total-energy} accurately describes dynamics of a transversal domain wall in thin curvilinear wire even for a magnetically soft ($K=0$) material: for plain wires \cite{Yershov15b} and microhelix wires \cite{Pylypovskyi15e}. Competition of the exchange and anisotropy contributions results in the length scale of the system $\Delta_0=\ell/\sqrt{k_{\text{t}}}$. For magnetically soft wires one has $\Delta_0=2\ell$.  In \eqref{eq:total-energy} and everywhere below we restrict ourselves with the case of a thin wire $\rho_0\lesssim \Delta_0$, this justifies our assumption of one-dimensionality of the problem: magnetization varies only along the wire and it is uniform within a wire cross-section. 

Up to now the curvilinearity formally does not appear in the problem \eqref{eq:LL-ZhLi}--\eqref{eq:total-energy}. However, the spin-polarized current and anisotropy axis are oriented tangentially to the wire. This makes it convenient to describe the magnetization distributions (e.g. domain walls) in terms of the local curvilinear basis:
\begin{equation} \label{eq:m-parameterization}
\begin{split}
\vec{m}=&m_\textsc{t}\vec{e}_\textsc{t}+m_\textsc{n}\vec{e}_\textsc{n} + m_\textsc{b}\vec{e}_\textsc{b}\\
=&\cos\theta\vec{e}_\textsc{t}+\sin\theta\cos\phi\vec{e}_\textsc{n}+\sin\theta\sin\phi\vec{e}_\textsc{b},
\end{split}
\end{equation}
where an angular representation is introduced to take into account the constraint $|\vec{m}|=1$. Here the crucial point is that the curvilinear basis vectors ($\vec{e}_\textsc{{t}},\,\vec{e}_\textsc{{n}},\vec{e}_\textsc{{b}}$) depend on spatial coordinate $s$. Therefore, all terms in \eqref{eq:LL-ZhLi} containing spatial derivatives introduce curvature $\kappa$ and torsion $\tau$ into the problem via Frenet--Serret formulas \eqref{eq:FS-formulas}. One can treat it as an emergence of new geometrically induced interactions. Thus, the exchange energy density $\mathscr{E}_{\text{ex}}$ in terms of the angular representation has the following form \cite{Sheka15}
\begin{equation} \label{eq:Eex-density}
\begin{split}
&\mathscr{E}_{\text{ex}} =\mathfrak{A}^2+\mathfrak{B}^2,\qquad\mathfrak{A}=\theta'+\kappa\cos\phi,\\
&\mathfrak{B}=\sin\theta(\phi'+\tau)-\kappa\cos\theta\sin\phi.
\end{split}
\end{equation}
In $\mathscr{E}_{\text{ex}}$, terms linear and bilinear with respect to $\kappa$ and $\tau$ can be treated as effective DMI and anisotropy interactions, respectively \cite{Sheka15}. Spin torques in \eqref{eq:LL-ZhLi} contain space derivatives $\left(\vec{u}\cdot\nabla\right)\vec{m}=u\,\vec{m}'$. Substituting the angular parameterization \eqref{eq:m-parameterization} into \eqref{eq:LL-ZhLi} and taking into account Frenet--Serret formulas \eqref{eq:FS-formulas} one obtains the following angular form of the equations of motion
\begin{equation} \label{eq:LL-ZhLi-angles}
\begin{split}
-\sin\theta\left(\dot{\theta}+u\theta'\right)=\omega_0\frac{\delta\mathcal{E}}{\delta\phi}+&u\kappa\sin\theta\cos\phi\\
+&\alpha\sin^2\theta\dot{\phi}+u\beta\sin\theta\,\mathfrak{B},\\
\sin\theta\left(\dot{\phi}+u\phi'\right)=\omega_0\frac{\delta\mathcal{E}}{\delta\theta}+&u\left(\kappa\cos\theta\sin\phi-\tau\sin\theta\right)\\
+&\alpha\dot{\theta}+u\beta\,\mathfrak{A}.
\end{split}	
\end{equation}
Note the correspondence between the nonadiabatic terms in \eqref{eq:LL-ZhLi-angles} and summands in the exchange energy density \eqref{eq:Eex-density}.

 One can easily verify that the equations \eqref{eq:LL-ZhLi-angles} are Lagrange--Rayleigh equations
\begin{equation*} 
\frac{\delta\mathcal{L}}{\delta\xi_i}-\frac{\mathrm{d}}{\mathrm{d}t}\frac{\delta\mathcal{L}}{\delta\dot\xi_i}=\frac{\delta\mathcal{F}}{\delta\dot\xi_i},\qquad\xi_i\in\{\theta,\,\phi\}
\end{equation*}
for the Lagrange function
\begin{subequations} \label{eq:Lagrange-Diss}
\begin{align} \label{eq:Lagrange-function}		&\mathcal{L}=-\mathcal{S}\int\limits_{-\infty}^{\infty}\phi\sin\theta(\dot\theta+u\theta')\mathrm{d}s-\omega_0\mathcal{E}-\mathcal{E}^{u},\\ 
\label{eq:energy}
&\mathcal{E}^{u}=u\mathcal{S}\int\limits_{-\infty}^{\infty}\left(\kappa\sin\theta\sin\phi+\tau\cos\theta\right)\mathrm{d}s,
\end{align}	
and dissipative function $\mathcal{F}=\mathcal{F}^{\mathrm{G}}+\mathcal{F}^u$, consisting of two summands: the ``standard'' Gilbert dissipative function~\cite{Gilbert04,Thiaville02a} 

\begin{align} \label{eq:Fdiss-Gilbert}
\mathcal{F}^{\mathrm{G}}=\frac{\alpha}{2}\mathcal{S}\int\limits_{-\infty}^{\infty}\left(\dot{\theta}^2+\sin^2\theta\dot{\phi}^2\right)\mathrm{d}s
\shortintertext{and ``nonadiabatic'' correction}
\label{eq:Dissipative-function}	\mathcal{F}^u=u\beta\mathcal{S}\int_{-\infty}^{\infty}\left(\mathfrak{A}\,\dot{\theta}+\mathfrak{B}\sin\theta\dot{\phi}\right)\mathrm{d}s.
\end{align}
\end{subequations}

Treatment of  the Landau--Lifshitz equation in terms of the Lagrange formalism was initially proposed by D\"{o}ring~\cite{Doering48}. In order to take into account the adiabatic spin-torque ($\beta=0$) the corresponding modification of the Lagrange function was proposed by Thiaville \cite{Thiaville04}, see the second summand in the integrand in \eqref{eq:Lagrange-function}. In this paper, in order to take into account the curvilinear effects and nonadiabaticity, we modify energy functional and dissipative function in accordance to \eqref{eq:energy} and \eqref{eq:Dissipative-function}, respectively.

\section{Collective variables approach}\label{sec:collective-variables}
To analyze the domain wall properties we use collective variable approach \cite{Slonczewski72,Thiele73,Malozemoff79} based on generalized $q$-$\Phi$ model~\cite{Kravchuk14}
\begin{equation} \label{eq:q-Phi-gen}
\theta=2\arctan e^{p(s-q)/\Delta},\qquad \phi=\Phi+a\frac{s-q}{\Delta}.
\end{equation}
Here $\left\{q,\, \Phi \right\}$ and $\left\{\Delta,\,a\right\}$ are pairs of time dependent conjugated collective variables: $q$ and $\Phi$ determine the domain wall position and phase (momentum), respectively; $\Delta$ and $a$ determine domain wall width and asymmetry of the phase distribution, respectively. Topological charge $p$ determines the domain wall type: head-to-head ($p=+1$) or tail-to-tail ($p=-1$).

Strictly speaking, the Ansatz \eqref{eq:q-Phi-gen} does not correspond to the ground states with distance from the domain wall position $|s-q|\gg\Delta$. Indeed, according to Ref.~\onlinecite{Sheka15c} the ground state of a helix wire with easy-tangential anisotropy can not be strictly tangential: the magnetization vector $\vec m$ deviates from the tangential direction by an angle $\psi\approx\kappa\tau \Delta_0^2$ (for small curvature and torsion). This can be treated as a result of action of an effective geometry induced magnetic field \cite{Sheka15}, which in case of a helix is oriented along binormal vector $\vec{e}_\textsc{{b}}$. Thus, in order to use the Ansatz \eqref{eq:q-Phi-gen} one has first rotate the basis ($\vec{e}_\textsc{{t}},\,\vec{e}_\textsc{{n}},\vec{e}_\textsc{{b}}$) by an angle $\psi$ around the normal vector $\vec{e}_\textsc{{n}}$ \cite{Pylypovskyi15e}. However, introducing the angular variables \eqref{eq:m-parameterization} in the rotated reference frame one obtains a set of equations of motion which coincide with \eqref{eq:LL-ZhLi-angles} up to corrections infinitesimal in $\kappa$ and $\tau$ of the second and third order for the energy $\mathcal{E}$ and spin-torque terms, respectively. Therefore in order to describe effects linear in $\kappa$ and $\tau$ the procedure of the basis adjustment \cite{Pylypovskyi15e} is not required. In the following we restrict ourselves to the linear in $\kappa$ and $\tau$ analysis in order to keep intelligibility of the solutions structure.

\subsection{Static solution}

Let us first consider the no driving case. Substituting Ansatz \eqref{eq:q-Phi-gen} into energy functional \eqref{eq:total-energy} and taking into account \eqref{eq:m-parameterization} and \eqref{eq:Eex-density} one obtains the following expression for the energy of a static domain wall 
\begin{equation} \label{eq:static-energy}
\frac{\mathcal{E}}{2\mathcal{S}}\approx\frac{\ell^2}{\Delta}(1+a^2)+k_t\Delta+2\ell^2\tau a+\ell^2\pi p \kappa \cos\Phi.
\end{equation}
The energy approximation \eqref{eq:static-energy} saves only  linear in $\kappa$ and $\tau$ terms in the corresponding equations for the variational parameters, whose equilibrium values read
\begin{equation} \label{eq:static-params}
\Delta_0\approx\ell/\sqrt{k_t},\qquad a_0\approx -\tau\Delta_0,\qquad\cos\Phi_0=-p.
\end{equation}
As well as for the case of a planar curvilinear wire \cite{Yershov15b} the \textit{curvature} fixes the domain wall magnetization in direction $-p\,\vec{e}_\textsc{{n}}$, thus the static head-to-head (tail-to-tail) domain wall is always magnetized outward (inward) the helix, see also Fig.~\ref{fig:stat-state}. This is due to the last term in \eqref{eq:static-energy}.  However, in contrast to the planar case, the new \textit{torsion} induced parameter of the domain wall asymmetry $a_0$ appears for the three dimensional wire. It confirms the recent results of Ref.~\onlinecite{Pylypovskyi15e} and it is similar to the domain wall asymmetry appearing due to the direct DMI \cite{Kravchuk14}. The domain walls structure which is determined by Ansatz \eqref{eq:q-Phi-gen} with parameters \eqref{eq:static-params} is compared with structure of domain walls obtained by means of micromagnetic \texttt{NMAG} simulations~ \cite{Fischbacher07}, see Fig.~\ref{fig:stat-state}. We consider four cases of different domain wall charge $p=\pm1$ and helix chirality $\mathcal{C}=\pm1$. In all cases the model \eqref{eq:q-Phi-gen} demonstrates a very good agreement with the simulations data. It should be noted that we perform a full scale micromagnetic simulations with two interactions taken into account, namely exchange and magnetostatic interactions. The obtained  conformity with the theory confirms the physical soundness of the model \eqref{eq:total-energy} which assume that the magnetostatic interaction can be reduced to the easy-tangential anisotropy in thin nanowires.

In simulations we use material parameters of Permalloy, namely: exchange constant $A = 13$ pJ/m, saturation magnetization $M_s = 860$ kA/m, damping coefficient $\alpha = 0.01$. These parameters result in the exchange length $\ell\approx3.7$ nm and $\omega_0 = 30.3$ GHz. Thermal effects and anisotropy are neglected. An irregular tetrahedral mesh with cell size about 2.75 nm is used.

\subsection{Domain wall dynamics}\label{sec:DW-dynamics}
\begin{figure*}
\includegraphics[width=\textwidth]{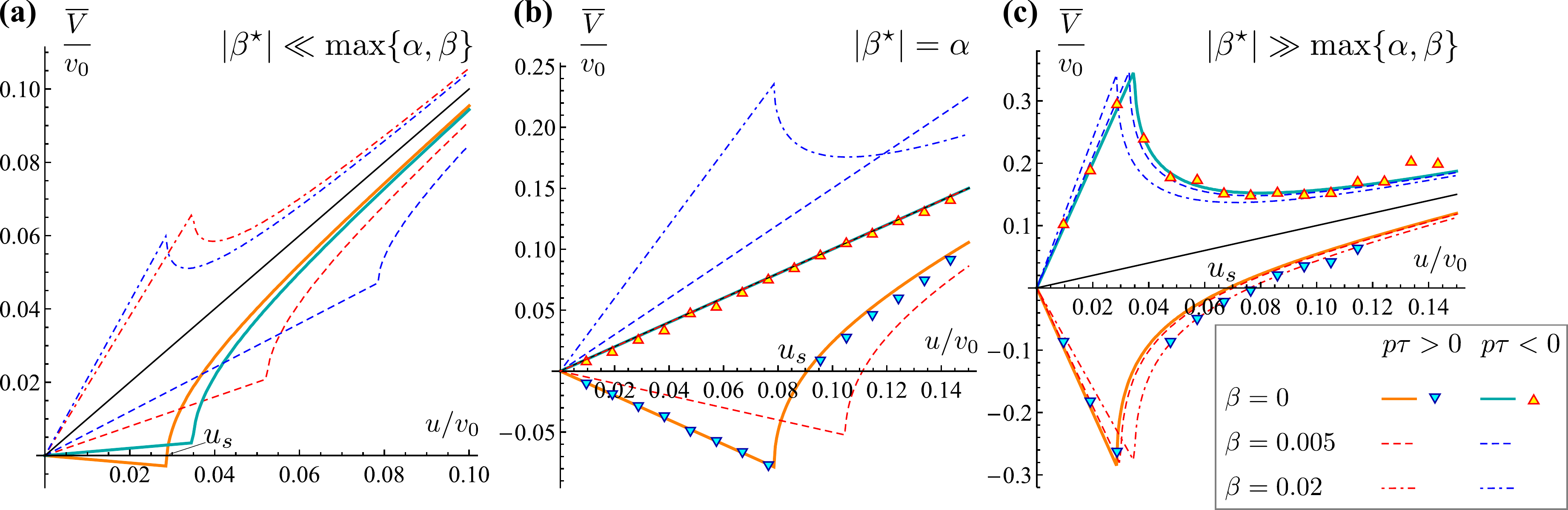}
\caption{\label{fig:Walker-plots}%
(Color online) Averaged domain wall velocity as a function of the applied current. Lines correspond to solutions of the collective variables equations \eqref{eq:q-Phi-main}, and markers show the results of \texttt{NMAG} micromagnetic simulations. Simulations as well as theoretical solutions are made for magnetically soft wire ($k_t=1/4$). All the other parameters are as follows: (a) $\kappa\ell=0.01$, $\tau\ell=0.0005$ ($|\beta^\star|=0.001$); (b) $\kappa\ell=0.05$, $\tau\ell=0.005$ ($|\beta^\star|=0.01$); (c) $\kappa\ell=0.1$, $\tau\ell=0.05$ ($|\beta^\star|=0.1$). In all cases $\alpha=0.01$. The thin solid line corresponds to the case of a rectilinear wire, where $V\approx u$.}
\end{figure*} 
Here we use a common collective variable approach which enables one to proceed from PDEs \eqref{eq:LL-ZhLi-angles} to a set of ODEs \eqref{eq:coll-vars-eqs}  with respect to the set of collective variables $\left\{a(t),\Phi(t),\Delta(t),a(t)\right\}$, for details see Appendix~\ref{app:CV-equations}. The obtained set  \eqref{eq:coll-vars-eqs} is nonlinear one, and only numerical analysis is possible in general. However, similarly to the case of a rectilinear wire, the two well separated timescales can be distinguished in this system. Dynamics of the pair $\left\{\Delta,\,a\right\}$ is characterized by the typical frequency $\omega\approx2\omega_0k_t/c$, where $c=\pi^2/12$, while the upper estimate of typical frequency of the pair $\left\{q,\, \Phi \right\}$ is $\Omega=p(\beta-\beta^\star-\alpha)u/\Delta_0$, where
\begin{equation*}
\beta^\star=p\tau\Delta_0,
\end{equation*}
for details see Appendix~\ref{app:CV-equations}. In practice, the timescales separation condition $\omega\gg|\Omega|$, which can be reformulated as
\begin{equation} \label{eq:sep-condition}
\frac{u}{v_0}|\beta-\beta^\star-\alpha|\ll\frac{2}{c}\sqrt{k_t},
\end{equation}
 is well satisfied due to small values of $\alpha$, $\beta$ and $\beta^\star$. Here and everywhere below we assume that $u>0$ with no loss of generality, and the notation $v_0=\ell\omega_0$ is used. Thus, the dynamics of the pair $\left\{\Delta,\, a\right\}$ is much faster than the dynamics of the pair $\left\{q,\, \Phi \right\}$, which in this case can be described by a set of two equations
\begin{equation} \label{eq:q-Phi-main}
\begin{split}
&\dot{q}-\alpha p\Delta_0\dot{\Phi}=u-\pi v_0\ell\kappa\sin\Phi,\\
&\alpha p\dot{q}+\Delta_0\dot{\Phi}=up(\beta-\beta^\star),
\end{split}	
\end{equation}
where we assume that the damping and the nonadiabaticity are low: $\alpha^2\ll1$, $\alpha a\ll1$, and $\beta^2\ll1$ and also the curvature is assumed to be small $\kappa\ell\ll1$. The values of the fast variables are determined by the values of slow ones in the following way
\begin{equation} \label{eq:Delta-fast}
\begin{split}
&\Delta(t)=\Delta[\Phi(t)]\approx\frac{\ell}{\sqrt{k_t+\frac{u}{v_0}\frac{\pi}{2}\kappa\ell\sin\Phi}}\\
&a(t)=a[\Phi(t)]\approx-\Delta\tau-p\frac{\pi}{4}\frac{u}{v_0}\frac{\beta\kappa\ell}{k_t}\sin\Phi.
\end{split}
\end{equation}

Equations \eqref{eq:q-Phi-main} have a traveling wave solution $q=Vt$ and $\Phi=\mathrm{const}$, which exists for the case $u<|u_c|$, where
\begin{equation} \label{eq:uc}
u_c\approx v_0\frac{\pi\alpha\kappa\ell}{\alpha-\beta+\beta^\star}
\end{equation}
is the Walker limit. The corresponding domain wall velocity is
\begin{equation} \label{eq:V}
V\approx u\frac{\beta-\beta^\star}{\alpha}.
\end{equation}
The value of the constant domain wall phase $\Phi$ is determined by the equation $\sin\Phi=u/u_c$.

Let us estimate the effective mass of he domain wall~\cite{Doering48}. To this end we consider a no driving case ($u=0$) with vanishing damping. In this case a small deviation $\tilde{\Phi}=\Phi-\Phi_0$ of the domain wall phase from its equilibrium value results in the traveling-wave domain wall motion with the velocity $V\approx p\pi v_0\kappa\ell\tilde{\Phi}$. Using the latter relation and energy expression \eqref{eq:static-energy} one can estimate energy of the moving domain wall as $E/\mathcal{S}\approx\varepsilon_0+\mathcal{M}V^2/2$, where $\varepsilon_0$ is energy density of a stationary domain wall and quantity
  \begin{equation} \label{eq:M}
  \mathcal{M}\approx\frac{8M_s^2}{\kappa v_0^2}
  \end{equation}
can be interpreted as an effective mass of the domain wall per unit area. For a permalloy helix with $\mathcal{R}=\mathcal{P}=1$~$\mu$m one obtains $\mathcal{M}\approx1.2\times10^{-24}$~kg/nm$^2$. This value is close one obtained experimentally for permalloy nanostripes~\cite{Saitoh04}. An infinite mass which one obtains in the limit case $\kappa\to0$ corresponds to the case of a rectilinear biaxial wire with vanishing transversal anisotropy \cite{Doering48,Tatara04,Thomas06,Kravchuk14}.  However, in contrast to the case of a rectilinear biaxial magnet, the domain wall mass \eqref{eq:M} does not depend on the longitudinal anisotropy $k_t$. This is because the mass \eqref{eq:M} originates mainly from the effective DMI term $\ell^2\pi p \kappa \cos\Phi$ in \eqref{eq:static-energy}, but not from the anisotropy contribution.

When the applied current achieves the critical value $u=|u_c|$ the system experiences a saddle node bifurcation, see Appendix~\ref{app:bifurcation} for details.  When the driving exceeds the critical value $u>|u_c|$ the domain wall demonstrates an precession motion with frequency 
\begin{equation} \label{eq:freq-osc}
\Omega_{\mathrm{prec}}=\Omega\sqrt{1-u^2/u_c^2},
\end{equation}
for details see Appendix~\ref{app:bifurcation}. This behavior is typical for the Walker limit overcoming. In the precession regime the domain wall can be characterized by some averaged in time drift velocity $\overline{V}$.  Fig.~\ref{fig:Walker-plots} demonstrates dependences $\overline{V}(u)$ for various values of parameters. One can see a typical domain wall behavior with the Walker limit being present. However a few special features should be marked out. A domain wall always moves in the direction of electrons flow ($\overline{V}>0$) for the case $p\tau<0$, while for the case $p\tau>0$ it can move in opposite direction. The latter case has two peculiarities: (i) if $\beta\ll|\beta^\star|$ then the traveling wave motion ($u<u_c$) is always characterized by negative mobility with $V<0$, (ii) for the precession regime ($u>u_c$) there always is a current value $u_s>u_c$ which corresponds to $\overline{V}=0$, i.e. the domain wall oscillates around some fixed position. It is important to emphasize that influence of torsion on the domain wall behavior is determined by the product $p\tau$, so the head-to-head and tail-to-tail domain walls swap roles when the helix chirality (sign of $\tau$) is changed to opposite one. It also should be noted that influence of the non-adiabatic parameter $\beta$ vanishes for the case $|\beta^\star|\gg\max\{\alpha,\,\beta\}$, see Fig.~\ref{fig:Walker-plots}.

\begin{figure}
	\includegraphics[width=0.8\columnwidth]{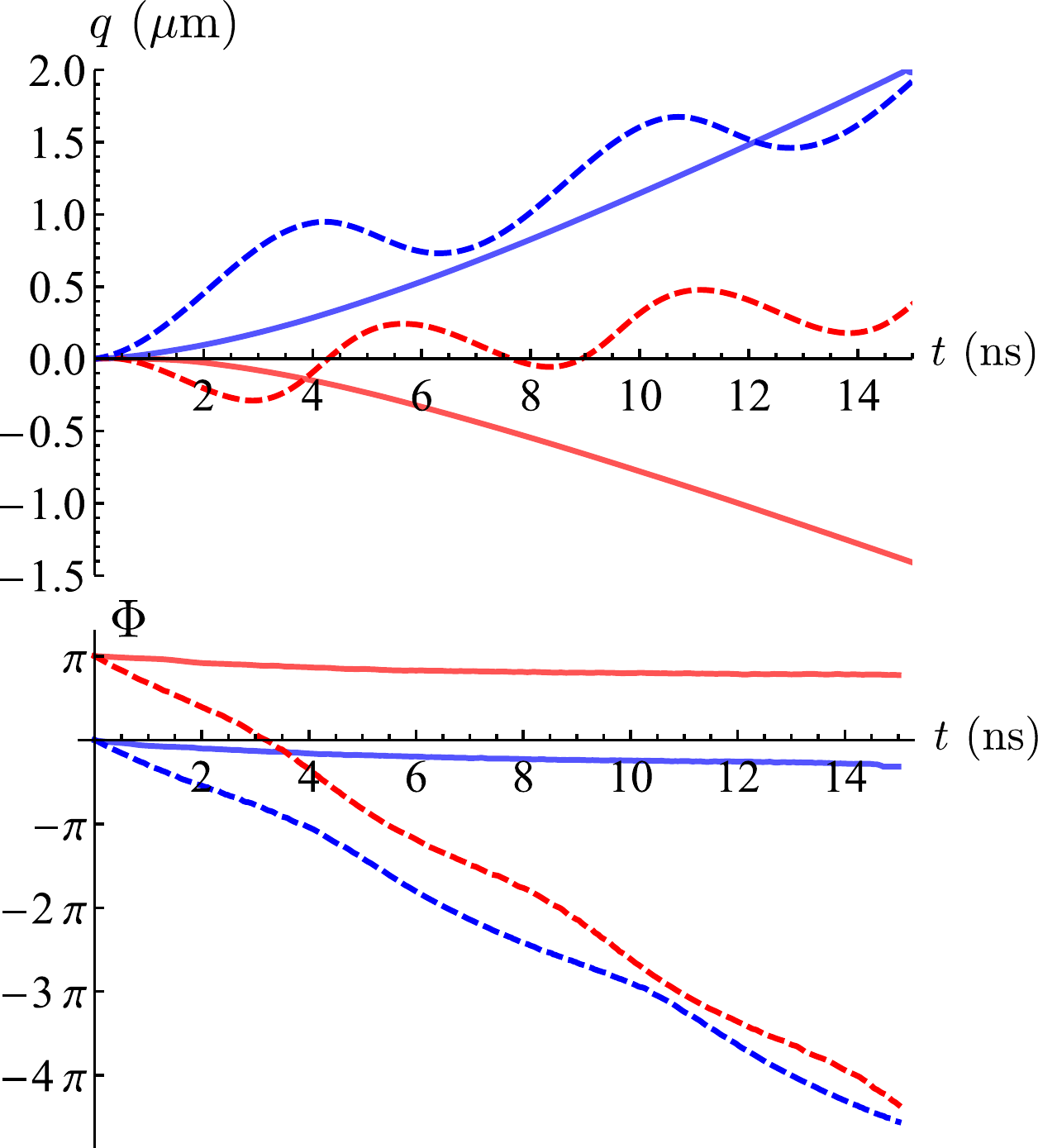}
	\caption{\label{fig:q-phi} %
	(Color online) Four examples of domain wall dynamics, which correspond to Fig.~\ref{fig:Walker-plots}(c). Solid and dashed lines correspond to the traveling-wave ($u<|u_c|$) and precession ($u>|u_c|$) regime, respectively, the corresponding dynamics is illustrated in the supplemental Video~\ref{mov:steady} and Video~\ref{mov:precession}. The data are obtained by means micromagnetic simulations.}
\end{figure}

Some examples of possible types of the domain wall motions are shown in the Fig.~\ref{fig:q-phi}. One can conclude that the domain wall dynamics in traveling-wave and precession regimes is similar to the corresponding domain wall dynamics in a biaxial magnet.

Finally, we would like to note how the result \eqref{eq:V} can be used for an experimental determining of parameters $\alpha$ and $\beta$. For this purpose one should measure the domain wall mobility $\mu=V/u$ for two helices with different torsions $\tau_1$ and $\tau_2$ and small curvatures. Then by using \eqref{eq:V} and taking into account that $\beta^\star=p\Delta_0\tau$, one obtains the following values of the parameters 
\begin{equation} \label{eq:alpha-beta}
	\alpha=p\Delta_0\frac{\tau_2-\tau_1}{\mu_1-\mu_2},\qquad \beta=p\Delta_0\frac{\mu_1\tau_2-\mu_2\tau_1}{\mu_1-\mu_2},
\end{equation}
where $\mu_i$ is the domain wall mobility in the helix with torsion $\tau_i$.

\section{Conclusions}\label{sec:concl}
The influence of curvature and torsion on the spin-current driven domain wall motion is studied by the example of helical wires. The analytical results are well confirmed by the full scale micromagnetic simulations. It is shown that the curvature results in the Walker limit appearance \eqref{eq:uc}, while the torsion effectively shifts the material parameter of nonadiabaticity $\beta\to\beta-\beta^\star$, where $\beta^\star=p\tau\Delta_0$, see \eqref{eq:q-Phi-main} and the consequences. This effect can lead to  a negative effective nonadiabaticity resulting in negative domain wall mobility. Significant influence of the wire torsion on the domain wall motion can be used for an experimental determination of the nonadiabatic parameter $\beta$ and damping coefficient $\alpha$, see \eqref{eq:alpha-beta}.

Additionally, we show that the effective mass of the domain wall is inversely proportional to the wire curvature. We demonstrate that the Walker limit is a saddle-node bifurcation. 
\appendix
\section{Equations of motion for collective variables}\label{app:CV-equations}
 Substituting the Ansatz \eqref{eq:q-Phi-gen} into \eqref{eq:Lagrange-Diss} and performing integration over $s$, one obtains the following effective Lagrange function
 \begin{subequations} \label{eq:L-CV}
\begin{equation} \label{eq:L-CV-1}
\mathcal{L}=2\mathcal{S}p\left[\Phi(\dot q-u)+c\dot{\Delta}a\right]-\omega_0\mathcal{E}-\mathcal{E}^u,
\end{equation}
where $c=\pi^2/12$, energy $\mathcal{E}$ is determined by \eqref{eq:static-energy} and the effective curvature induced spin-torque correction of the energy reads
\begin{equation} \label{eq:Eu-CV}
\mathcal{E}^u\approx \mathcal{S}u\left(\pi\kappa\Delta\sin\Phi+2p\tau q\right).
\end{equation}
\end{subequations}
As it follows from \eqref{eq:L-CV}, the domain wall asymmetry parameter $a$ is an conjugated momentum to the domain wall width $\Delta$. In the same way one obtains components of the effective dissipative function $\mathcal{F}=\mathcal{F}^{\mathrm{G}}+\mathcal{F}^u$: the Gilbert part 
 \begin{subequations} \label{eq:F-CV}
\begin{equation} \label{eq:F-Gilbert-CV}
\mathcal{F}^{\mathrm{G}}=\frac{\alpha}{\Delta}\mathcal{S}\left\{\dot{q}^2+(\dot{q}a-\dot{\Phi}\Delta)^2+c\left[\dot{\Delta}^2+(\dot{a}\Delta-\dot{\Delta}a)^2\right]\right\}
\end{equation}
coincides with previously obtained one \cite{Kravchuk14} and the nonadiabatic correction reads 
\begin{equation} \label{eq:F-beta-CV}
\begin{split}
\mathcal{F}^u\approx-2u\beta\mathcal{S}\biggl[&\frac{\dot{q}}{\Delta}-\dot{\Phi}(a+\Delta\tau)\\ \nonumber
&+p\frac{\pi}{2}\kappa\left(\dot{q}\cos\Phi-\dot{a}\Delta\sin\Phi\right)\biggr].
\end{split}
\end{equation}
\end{subequations}
Lagrange function \eqref{eq:L-CV} and dissipative function \eqref{eq:F-CV} generate the following set of equations of motion for the collective variables
\begin{subequations} \label{eq:coll-vars-eqs}
\begin{align} \label{eq:q-eq}
&(p+\alpha a)\dot{q} - \alpha\Delta\dot{\Phi} = pu\left(1+p\frac{\pi}{2}\kappa\Delta\cos\Phi\right)\\ \nonumber 
&-p\pi v_0\ell\kappa\sin\Phi+u\beta(a+\Delta\tau),\\
\label{eq:Phi-eq}
&\alpha\frac{\dot{q}}{\Delta}+(p-\alpha a)\dot{\Phi} = \frac{u\beta}{\Delta}\left(1+p\frac{\pi}{2}\kappa\Delta\cos\Phi\right)-pu\tau,\\
\label{eq:a-eq}	
&\frac{c}{\omega_0}\!\left[(p+\alpha a)\frac{\dot{\Delta}}{\Delta} - \alpha\dot{a}\right]\!\!=2\frac{\ell^2}{\Delta^2}\left(a+\Delta\tau\right)+p\frac{\pi}{2}\frac{u}{v_0}\beta\kappa\ell\sin\Phi,\\
\label{eq:Delta-eq}
&\frac{c}{\omega_0}\!\left[\alpha\frac{\dot{\Delta}}{\Delta}+(p-\alpha a)\dot{a}\right]\!\!=\frac{\ell^2}{\Delta^2}(1+a^2)-k_t-\frac{\pi}{2}\frac{u}{v_0}\kappa\ell\sin\Phi,
\end{align}
\end{subequations}
where $v_0=\ell\omega_0$.

First of all, it should be noted that in the no driving case ($u=0$) equations \eqref{eq:a-eq} and \eqref{eq:Delta-eq} split off the whole set \eqref{eq:coll-vars-eqs} forming an independent set of equations with respect to $\Delta$ and $a$. The linearization in vicinity of the stationary point $\left\{\Delta_0,\, a_0\right\}$ with respect to small deviations $\tilde{\Delta}=\Delta-\Delta_0$ and $\tilde{a}=a-a_0$ results in a set of linear equations
\begin{equation} \label{eq:Delta-a-linear}
\begin{Vmatrix}
\dot{\tilde{\Delta}}/\Delta_0\\
\dot{\tilde{a}}
\end{Vmatrix}\approx-\frac{2\omega_0}{c}\frac{\ell^2}{\Delta_0^2}
\begin{Vmatrix}
\alpha-\beta^\star&-p\\
p&\alpha+\beta^\star
\end{Vmatrix}\cdot
\begin{Vmatrix}
\tilde{\Delta}/\Delta_0\\
\tilde{a}
\end{Vmatrix},
\end{equation}
where $\beta^\star=p\Delta_0\tau$. In \eqref{eq:Delta-a-linear} the stationary values $\Delta_0$ and $a_0$ are determined by \eqref{eq:static-params} and the low damping approximation ($\alpha^2\ll1$, $\alpha a_0\ll1$) is applied. Equations \eqref{eq:Delta-a-linear} have a solution $\tilde{\Delta}=\tilde{\Delta}_0\exp(-\eta t+i\omega t)$ and $\tilde{a}=\tilde{a}_0\exp(-\eta t+i\omega t)$, where
\begin{equation} \label{eq:DW-osc-params}
\omega\approx\frac{2\omega_0}{c}k_t,\qquad \eta=\alpha\omega.
\end{equation}
Let us now assume that $\omega\gg\Omega$, where $\Omega$ is characteristic frequency of the pair $\left\{q,\, \Phi \right\}$, i.e. pair $\left\{\Delta,\,a \right\}$ is much faster than the pair $\left\{q,\,\Phi \right\}$. In this case the quasi-stationary values of $\Delta$ and $a$ are slightly modified, see \eqref{eq:Delta-fast}. 

However the time characteristics of the pair $\left\{\Delta,\,a \right\}$ are the same as \eqref{eq:DW-osc-params}. Taking into account \eqref{eq:Delta-fast} one can split off the first two equations of the set \eqref{eq:coll-vars-eqs} and present them in the form \eqref{eq:q-Phi-main}.

The characteristic frequency $\Omega$ can be easily determined in the case of small curvature. Indeed in the limit case of vanishing curvature $\kappa\to0$ and low damping a couple of equations \eqref{eq:q-eq} and \eqref{eq:Phi-eq} takes a form
\begin{equation} \label{eq:qPhi-no-curvature}
p\dot{q}-\alpha\Delta_0\dot{\Phi}=pu,\qquad \alpha\dot{q} + p\Delta_0\dot{\Phi} = u(\beta-\beta^\star),
\end{equation}
which coincide with the well known \cite{Thiaville05} $q$-$\Phi$ equations for the spin-current driven domain wall motion in a rectilinear uniaxial wire, except shift of the nonadiabatic parameter $\beta\to\beta-\beta^\star$. Equations \eqref{eq:qPhi-no-curvature} have solution in form of uniform domain wall motion $q=Vt$ with uniform phase precession $\Phi=\Omega t$, where 
\begin{equation} \label{eq:V-Omega}
V\approx u,\qquad \Omega\approx p\frac{u}{\Delta_0}(\beta-\beta^\star-\alpha).
\end{equation}
In order to take the curvature into account we consider it as a small perturbation, which results in small deviations of the collective variables: $q=Vt+\tilde q$ and $\Phi=\Omega t+\tilde{\Phi}$. Substituting it in \eqref{eq:q-Phi-main} and taking into account \eqref{eq:V-Omega} one obtains
\begin{equation*} 
\begin{split}
&\dot{\tilde{q}}=\pi v_0\kappa\ell\sqrt{1+\frac{u^2}{v_0^2}\frac{1}{4k_t}}\cos\left(\Omega t+\delta_q\right),\\
&\dot{\tilde{\Phi}}=\pi \omega_0\kappa\ell\sqrt{\alpha^2k_t+\frac{1}{4}\frac{u^2}{v_0^2}(\beta-\alpha)^2}\sin\left(\Omega t+\delta_\Phi\right),\\
&\tan\delta_q=2\sqrt{k_t}p\frac{v_0}{u},\quad\tan\delta_\Phi=p\frac{u}{v_0}\frac{\beta-\alpha}{\alpha}\frac{1}{2\sqrt{k_t}}.
\end{split}
\end{equation*}
Thus, the curvature add an oscillatory component of frequency $\Omega$ to the uniform domain wall motion with velocity $V$. The presented analysis enable one to make an upper estimate of the frequency $\Omega$ (for the case of high current $u$). So the condition of separation of time scales in \eqref{eq:coll-vars-eqs} reads $\omega\gg\Omega$, which can be also written as \eqref{eq:sep-condition}.

\section{Bifurcational analysis of equations of motion}
\label{app:bifurcation}

\begin{figure*}
	\includegraphics[width=\textwidth]{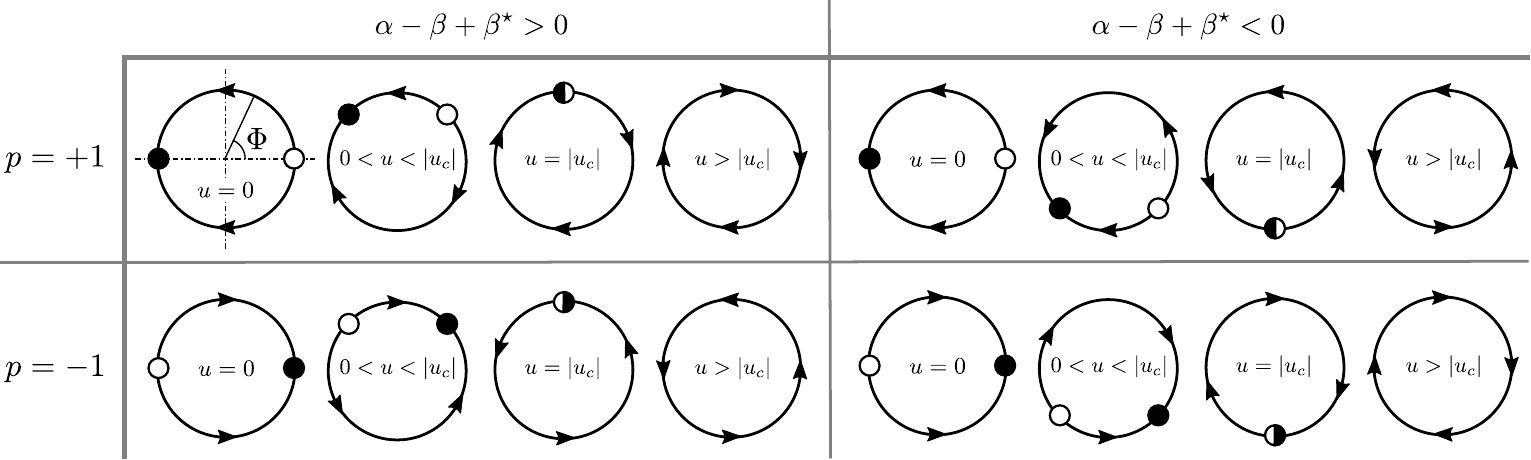}
	\caption{\label{fig:bifurcatoin}%
	All possible bifurcational diagrams for Eq.~\eqref{eq:oscillator} for various values of parameters. Filled $\CIRCLE$ and open $\Circle$ circles show stable and unstable fixed points, respectively.}
\end{figure*}

Let us now make a bifurcational analysis of the system \eqref{eq:q-Phi-main}. Excluding $\dot{q}$ from  \eqref{eq:q-Phi-main} one obtains (in a low damping limit) the equation
\begin{equation} \label{eq:oscillator}
\dot{\Phi}\approx\Omega\left(1-\frac{u_c}{u}\sin\Phi\right).
\end{equation} 
In mechanics, Eq.~\eqref{eq:oscillator} describes dynamics of an overdamped pendulum driven by a constant torque. For the case $u<|u_c|$ the equation \eqref{eq:oscillator} has two fixed points, namely the stable one
\begin{equation} \label{eq:Phi-stable}
\Phi_{\mathrm{s}}=\frac{1+p}{2}\pi-p\arcsin\frac{u}{u_c},
\end{equation}
see filled circles in the bifurcation diagram, Fig.~\ref{fig:bifurcatoin}, and unstable one
\begin{equation} \label{eq:Phi-unstable}
\Phi_{\mathrm{a}}=\frac{1-p}{2}\pi+p\arcsin\frac{u}{u_c},
\end{equation}  
see open circles in the bifurcation diagram, Fig.~\ref{fig:bifurcatoin}. In general, criterion of stability of the fixed point $\Phi^*\in\{\Phi_{\mathrm{s}},\Phi_{\mathrm{a}}\}$ reads $p\cos\Phi^*<0$. One has the only one stable fixed point, in contrast to the biaxial magnet, where two stable fixed points are present. 

Let us now consider behavior of the system at vicinity of the bifurcation point $u=|u_c|$. As it follows from \eqref{eq:Phi-stable}, $\Phi_{\mathrm{s}} \to \mathrm{sgn}(\alpha - \beta + \beta^\star) \pi/2$ when $u\to |u_c|$. Introducing now small deviations $\tilde\Phi=\Phi-\Phi_{\mathrm{s}}$ and $\tilde{u}=u-|u_c|$, one can easily obtain from \eqref{eq:oscillator}
\begin{equation} \label{eq:normal-form}
\Delta_0\dot{\tilde{\Phi}}=-p(\alpha-\beta+\beta^\star)\left(\tilde{u}+|u_c|\tilde{\Phi}^2/2\right).
\end{equation}
Relation \eqref{eq:normal-form} is a normal form for a saddle-node bifurcation \cite{Strogatz94}. All possible scenarios of the bifurcation are collected in the Fig.~\ref{fig:bifurcatoin}.
	
When the applied current exceeds the critical value $u>|u_c|$ the domain wall demonstrates a precession motion which is typical for the Walker limit overcoming. Period of the precession can be easily obtained by integrating \eqref{eq:oscillator}:
\begin{equation} \label{eq:period}
T=\frac{2\pi}{\Omega}\frac{1}{\sqrt{1-u_c^2/u^2}}.
\end{equation}
This corresponds to the frequency \eqref{eq:freq-osc}.
For the case $u\gtrapprox|u_c|$ one can estimate period \eqref{eq:period} as follows  $T\approx\pi\Omega^{-1}\sqrt{2|u_c|}/\sqrt{u-|u_c|}$. The obtained square-root scaling law is a very general feature of systems that are close to a saddle-node bifurcation \cite{Strogatz94}.
	
\section{Supplemental movie}
\label{app:movie}

The magnetization dynamics is studied by means of numerical simulation of the Landau--Lifshitz equation with additional Bazaliy--Zhang--Li spintorque terms applying the \texttt{NMAG} code~ \cite{Fischbacher07}. Only two magnetic interactions were taken into account, namely exchange and magnetostatic interactions. 

Simulation parameters: exchange constant $A = 13$ pJ/m, saturation magnetization $M_s = 860$ kA/m, damping coefficient $\alpha = 0.01$, nonadiabatic spin-transfer parameter $\beta=0$. Thermal effects and anisotropy are neglected. An irregular tetrahedral mesh with cell size about 2.75 nm is used.

Wire parameters: length 4 $\mu$m, radius 5 nm, curvature $\kappa\ell=0.1$, torsion $\tau\ell=0.05$.

The typical examples of steady-state and precession motion of head-to-head and tail-to-tail domain walls under influence of spin-polarized current are presented in Video~\ref{mov:steady} and Video~\ref{mov:precession}, respectively.
\begin{video} 
	\includemedia[
	label=steady,
	activate=onclick,
	width=0.9\columnwidth,height=0.5125\columnwidth,
	flashvars={modestbranding=1 
	  & autohide=1 
	  & showinfo=0 
	  & rel=0 
	  }
	]{\includegraphics[width=\columnwidth]{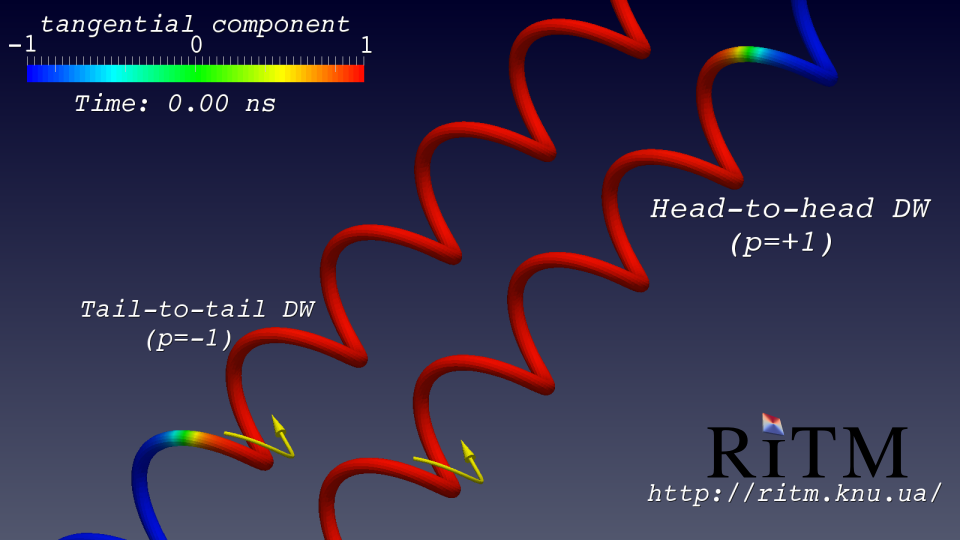}}{https://www.youtube.com/v/tTnc-XrdzWs?rel=0}
	\setfloatlink{https://www.youtube.com/watch?v=tTnc-XrdzWs}
	\caption{\label{mov:steady}%
	Steady-state motion of head-to-head and tail-to-tail domain walls under influence of spin-polarized current~($u/v_0=0.02$, $\beta=0$) in the helix nanowire. Yellow arrow determines the current direction~($\vec{u}=u\vec{e}_\textsc{t}$). The mobility for different types of domain walls has different sign, while the direction of current is the same.}
\end{video}

\begin{video} 
	\includemedia[
	label=precession,
	activate=onclick,
	width=0.9\columnwidth,height=0.5125\columnwidth,
	flashvars={modestbranding=1 
	  & autohide=1 
	  & showinfo=0 
	  & rel=0 
	  }
	]{\includegraphics[width=\columnwidth]{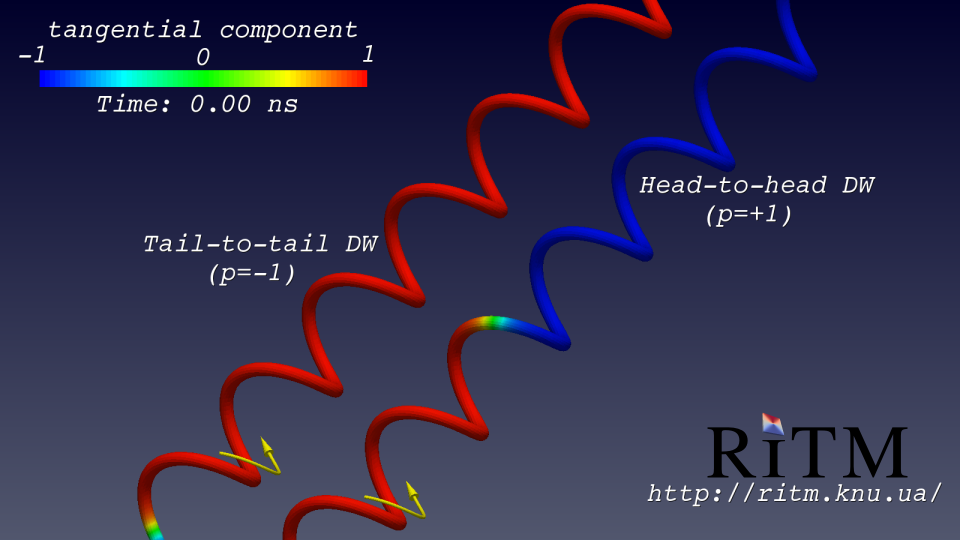}}{https://www.youtube.com/v/Zc7wgYA08cc?rel=0}
	\setfloatlink{https://www.youtube.com/watch?v=Zc7wgYA08cc}
	\caption{\label{mov:precession}%
	Precession motion of head-to-head and tail-to-tail domain walls under influence of spin-polarized current~($u/v_0=0.115$, $\beta=0$) in the helix nanowire. Yellow arrow determines the current direction~($\vec{u}=u\vec{e}_\textsc{t}$).}
\end{video}

%

%
\end{document}